\documentclass{elsart}
\usepackage{graphicx}
\usepackage{amssymb}

\def\lQ{\Lambda_{\rm QCD}}
\def\als{\alpha_{\rm s}} 
\def\siml{{\ \lower-1.2pt\vbox{\hbox{\rlap{$<$}\lower6pt\vbox{\hbox{$\sim$}}}}\ }}
\def\MS{\overline{\rm MS}}
\newcommand{\be}{\begin{equation}}
\newcommand{\ee}{\end{equation}}
\newcommand{\bea}{\begin{eqnarray}}
\newcommand{\eea}{\end{eqnarray}}
\newcommand{\nn}{\nonumber}

\begin{document}
\begin{frontmatter}
\begin{flushright}
\tt{ANL-HEP-PR-06-76 \\ IFUM-872-FT \\ UB-ECM-PF 06/27}
\end{flushright}

\title{The logarithmic contribution to the QCD static energy at N$^4$LO}
\author[Milano]{Nora Brambilla},
\author[ANL,BCN]{Xavier Garcia i Tormo},
\author[BCN]{Joan Soto},
\author[Milano]{Antonio Vairo}
\address[Milano]{Dipartimento di Fisica dell'Universit\`a di Milano and
  INFN, via Celoria 16, 20133 Milan, Italy}
\address[ANL]{High Energy Physics Division, Argonne National Laboratory, 9700 South Cass Avenue, Argonne, IL 60439, USA\thanksref{ara}}
\address[BCN]{Departament d'Estructura i Constituents de la Mat\`eria, 
     Universitat de Barcelona, Diagonal 647, E-08028 Barcelona, Catalonia, Spain}
\thanks[ara]{Current address}

\begin{abstract}
Using pNRQCD and known results for the field strength correlator, we calculate
the ultrasoft contribution to the QCD static energy of a quark-antiquark pair 
at short distances at N$^4$LO in $\als$. At the same order, this provides 
the logarithmic terms of the singlet static potential in pNRQCD and
the $\log \als$ terms of the static energy.
\end{abstract}
\end{frontmatter}

\section{Introduction}
The ground state energy, $E_0(r)$, of a static quark and a static antiquark
separated by a distance $r$ is a key object for the understanding of the QCD
dynamics. It is also a basic ingredient of the Schr\"odinger-like formulation of heavy quarkonium systems \cite{Brambilla:2004wf}.
Its linear behavior at long distances is a signal for confinement
\cite{Wilson:1974sk}, but also at short distances ($r \ll 1/\lQ$), where weak coupling
calculations are reliable, it shows a non-trivial behavior. Indeed, when
calculated in perturbation theory, infrared divergences are found
starting at three loops \cite{Appelquist:1977tw,Appelquist:1977es}. 
These are due to the virtual emission of ultrasoft gluons with energy of the
order $E_0(r)$, which turn a color singlet quark-antiquark pair into a color octet
one and vice-versa. The proper treatment of the ultrasoft emissions requires
the resummation of an infinite class of diagrams, which produces a
non-analytic dependence on $\als$ (typically logarithms of it).
We shall focus here on this short distance behavior. 

The current knowledge of $E_0(r)$ at short distance may be summarized as follows
\bea
E_0(r)&=&-\frac{C_F\als(1/r)}{r}\Bigg\{1+\frac{\als(1/r)}{4\pi}\left[a_1+2\gamma_E\beta_0\right]
\nn\\
&& +\left(\frac{\als(1/r)}{4\pi}\right)^2
\left[ a_2+ \left(\frac{\pi^2}{3}+4\gamma_E^2\right)\beta_0^2+\gamma_E\left(4a_1\beta_0+2\beta_1\right)\right]
\nn\\
&&
+\left(\frac{\als(1/r)}{4\pi}\right)^3\left[\frac{16\pi^2}{3}C_A^3\log{C_A
    \als(1/r)\over 2} + \tilde{a}_3\right]
\nn\\
&&
+\left(\frac{\als(1/r)}{4\pi}\right)^4
\left[ a_{4}^{L2}\log^2 {C_A \als(1/r)\over 2}+a_{4}^{L}\log {C_A
  \als(1/r)\over 2}+ \tilde{a}_{4} \right] 
\nn\\
&&
+ \qquad\qquad  \cdots \qquad\qquad\Bigg\},
\label{E0stat}
\eea
where $C_F= T_F (N_c^2-1)/N_c$, $C_A=N_c$, $T_F=1/2$, $N_c$ is the number of colors,
$\beta_0 = 11 C_A/3 - 4 T_F n_f/3$ and 
$\beta_1 = 34 C_A^2/3 - 20 C_A T_F n_f/3 - 4 C_F T_F n_f$ 
are the first two coefficients of the beta function,
$n_f$ is the number of (massless) flavors, $\gamma_E$ is the Euler constant
and $\als$ is the strong coupling constant in the $\MS$ scheme. 
The one-loop coefficient $a_1$ is given by \cite{Fischler:1977yf,Billoire:1979ih} 
\be
a_1=\frac{31}{9}C_A-\frac{20}{9}T_Fn_f, 
\ee
and the two loop coefficient $a_2$ by \cite{Peter:1996ig,Peter:1997me,Schroder:1998vy,Kniehl:2001ju}
\bea
a_2&=& \left({4343\over162}+4\pi^2-{\pi^4\over4}+{22\over3}\zeta(3)\right)C_A^2
-\left({1798\over81}+{56\over3}\zeta(3)\right)C_AT_Fn_f
\nn\\
&& -\left({55\over3}-16\zeta(3)\right)C_FT_Fn_f +\left({20\over9}T_Fn_f\right)^2.
\eea
The logarithmic piece of the third-order correction was calculated in
\cite{Brambilla:1999qa}, whereas the non-logarithmic piece $\tilde{a}_3$ has not been calculated yet. However, $\tilde{a}_3$ is believed to be dominated by contributions which are known from renormalization group arguments \cite{Chishtie:2001mf}. If we write $\tilde{a}_3={a}_3+ {a}_3^{\rm RG}\,$, $\,{a}_3^{\rm RG} \gg {a}_3\,$. $\,{a}_3^{\rm RG}$ has a known expression in terms of the coefficients of the beta function and of those entering in the potential at lower orders (see \cite{Chishtie:2001mf}, where $\tilde{a}_3=-48\pi^3\times V_3\,$, $\, {a}_3=64 c_0$). Estimates of ${a}_3$ have been carried out using Pad\'e approximations \cite{Chishtie:2001mf} and renormalon dominance \cite{Pineda:2001zq,Pineda:2002se,Cvetic:2003wk,Lee:2003hh}, which are consistent with the inequality above, and give similar results.
The double logarithmic fourth-order correction may be obtained from
\cite{Pineda:2000gz}, where higher-order contributions of the form
$\als^{n+3} \log^n \als $ were resummed using renormalization group techniques. It reads
\be
a_{4}^{L2}=\frac{16\pi^2}{3}C_A^3\left(-\frac{11}{3}C_A+\frac{4}{3}T_Fn_f\right).
\label{a4L2}
\ee

The main result of this letter is the calculation of the logarithmic 
fourth-order correction to the singlet potential,  $a_{4}^{L}$; we anticipate
that it reads
\bea
a_4^L &=&  16\pi^2C_A^3\left[a_1+2\gamma_E\beta_0 
+ T_F n_f \left( -\frac{40}{27} + \frac{8}{9} \log 2\right)
\right.
\nn\\
&&\qquad\qquad\qquad\qquad\qquad\qquad\qquad
\left.
+ C_A\left(\frac{149}{27}-\frac{22}{9}\log 2+\frac{4}{9}\pi^2\right)\right].
\label{a4L}
\eea
The non-logarithmic piece $\tilde{a}_4$ remains unknown.

A convenient method to calculate the logarithmic contributions to Eq. (\ref{E0stat}), 
which steam from the dynamics at the ultrasoft scale $E_0(r)$, consists in
integrating out from static QCD degrees of freedom at the soft energy scale $1/r$ and 
working within the effective field theory framework of pNRQCD 
\cite{Pineda:1997bj,Brambilla:1999xf}(see \cite{Brambilla:2004jw} for a review).
The quark-antiquark system may be in a color singlet or in a color octet
configuration, which are encoded in color singlet, S, and color octet, O, fields in pNRQCD.
At leading order in the multipole expansion, the integration of the soft
energy scale gives rise to a singlet, $V_s(r;\mu )$, and an octet, $V_o(r;\mu)$, static potential, which
depend on $r$ and a factorization scale $\mu$. At next-to-leading order,  
two more ``potentials'' appear, $V_A(r;\mu ) $ and $V_B(r;\mu )$, 
which are the matching coefficients of the singlet-octet and octet-octet vertices
respectively. At this order, the pNRQCD Lagrangian reads
\bea
{\mathcal L}_{\rm pNRQCD} &=& 
{\mathcal L}_{\rm light} 
\nn\\
&& 
+ \int d^3{\bf r} \; {\rm Tr} \,  
\Biggl\{ {\rm S}^\dagger \left[ i\partial_0 
- V_s(r;\mu ) \right] {\rm S} 
+ {\rm O}^\dagger \left[ iD_0 
- V_o(r;\mu ) \right] {\rm O} \Biggr\}
\nn\\
&& \qquad
+ V_A ( r; \mu) {\rm Tr} \left\{  {\rm O}^\dagger {\bf r} \cdot g{\bf E} \,{\rm S}
+ {\rm S}^\dagger {\bf r} \cdot g{\bf E} \,{\rm O} \right\} 
\nn\\
&& \qquad
+ {V_B (r; \mu) \over 2} {\rm Tr} \left\{  {\rm O}^\dagger {\bf r} \cdot g{\bf E} \, {\rm O} 
+ {\rm O}^\dagger {\rm O} {\bf r} \cdot g{\bf E}  \right\}
\nn\\
&& \qquad
+ \dots ,
\label{pNRQCD}
\eea
where ${\mathcal L}_{\rm light}$ is the part of the Lagrangian involving
gluons and light quarks, and coincides with the QCD one.
The dots stand for higher-order terms in the multipole expansion. 
The static energy calculated from the above Lagrangian has the form
\be
E_0(r)= V_s(r;\mu ) + \delta_{\rm US}(r,V_s,V_o,V_A,V_B, \dots;\mu),
\label{e0}
\ee
where $ \delta_{\rm US}(r,V_s,V_o,V_A,V_B, \dots;\mu)$ ($\delta_{\rm US}$ for short)
contains contributions from the ultrasoft gluons.
$V_s(r;\mu )$ and $V_o(r;\mu )$ do not depend on $\mu$ up to N$^2$LO \cite{Brambilla:1999xf}. 
The former coincides with $E_0(r)$ at this order and the latter may be found in \cite{Kniehl:2004rk}. 
The fact that the $\mu$ dependence of $\delta_{\rm US}$ must cancel 
the one in $V_s(r;\mu )$ is the key observation that 
leads to a drastic simplification in the calculation of the $\log \als$
terms in $E_0(r)$. So, for instance, the logarithmic contribution at N$^3$LO, 
which is part of the three-loop contributions to $V_s(r;\mu)$, may be extracted from a one-loop
calculation of $\delta_{\rm US}$ \cite{Brambilla:1999qa,Brambilla:1999xf} and the single logarithmic contribution at N$^4$LO,
which is part of the four-loop contributions to $V_s(r;\mu)$,
may be extracted from a two-loop calculation of $\delta_{\rm US}$.\footnote{
We denote N$^n$LO, contributions to the potential of order
$\als^{n+1}$ and N$^n$LL, contributions  of order $\als^{n+2} \log^{n-1}\als$.
}

In Sec. \ref{sec2}, we review the calculation of the third-order logarithmic
term since it follows the same lines as that of the fourth-order one, which will
be presented in Sec. \ref{sec3}. In Sec. \ref{sec4}, we conclude and discuss
some applications of this work.

\section{Review of the third-order logarithmic correction}
\label{sec2}
In $d$ dimensions, the order $r^2$ contribution due to ultrasoft gluons reads \cite{Brambilla:1999qa,Brambilla:1999xf}
\be
\label{GpNRQCD}
\delta_{\rm US} = -i{g^2 \over N_c}\,T_F\, V_A^2\,{r^2 \over d-1} \,\int_0^\infty dt \,
e^{-it(V_o-V_s)} \langle 0| {\bf E}^a(t) \phi(t,0)^{\rm adj}_{ab}{\bf E}^b(0) |0\rangle.
\ee
$ \phi(y,x)^{\rm adj}_{ab}$ is the Wilson line in the adjoint representation
connecting the points $y$ and $x$ by a straight line ($t$ stands for $(t,{\bf
  0})$). 
We will evaluate Eq. (\ref{GpNRQCD}) perturbatively in $\als$. 
The dependence on $\als$, apart from the trivial $g^2$ factor, enters through 
{\it (i)} the $V_s$ and $V_o$ potentials, 
{\it (ii)} $V_A$
and {\it (iii)} the field strength correlator of the chromoelectric fields.

{\it (i)} The difference $V_o-V_s$ is given at leading-order by
$\displaystyle\frac{C_A}{2}\frac{\als(1/r)}{r}$. Note that at leading and
next-to-leading order $V_s$ and $V_o$ only differ by an overall color factor.

{\it (ii)} At tree level $V_A = 1$.

\begin{figure}
\centering
\includegraphics{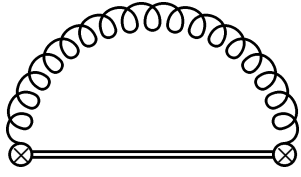}
\caption[Field strength correlator at leading order]{Leading-order
contribution to the field strength correlator. 
The double line represents the gluonic string, the circled cross 
and the springy line the chromoelectric field
correlator. Note that our convention differs from the one in
\cite{Eidemuller:1997bb}, where the gluonic string is either not represented, 
if no gluons emerge from it, or is represented by a dashed line.}
\label{figfscLO}
\end{figure}

{\it (iii)} The two-point field strength correlator
\be
\mathcal{D}_{\mu\nu\lambda\omega}(z)\equiv\left<0\right\vert T\left\{G_{\mu\nu}^a(y)
 \phi(y,x)^{\rm adj}_{ab}
G_{\lambda\omega}^b(x)\right\}\left\vert 0 \right>
\ee
can be parameterized in terms of two scalar functions $\mathcal{D}(z^2)$ and $\mathcal{D}_1(z^2)$ according to
\bea 
\mathcal{D}_{\mu\nu\lambda\omega}(z)&=&
\left(g_{\mu\lambda}g_{\nu\omega}-g_{\mu\omega}g_{\nu\lambda}\right)\left(\mathcal{D}(z^2)+\mathcal{D}_1(z^2)\right)
\nn\\
&& +\left(g_{\mu\lambda}z_\nu z_\omega-g_{\mu\omega}z_\nu z_\lambda
-g_{\nu\lambda} z_\mu z_\omega+g_{\nu\omega}z_\mu z_\lambda\right)
\frac{\partial\mathcal{D}_1(z^2)}{\partial z^2},
\eea
where $z=y-x$ \cite{Dosch:1988ha}. 
In (\ref{GpNRQCD}), $x$ and $y$ only differ in the time component, hence
$z=t$. Furthermore, in $d$ dimensions, the chromoelectric component is given by 
\bea
\langle 0| {\bf E}^a(y) \phi(y,x)^{\rm adj}_{ab}{\bf E}^b(x) |0\rangle
= \mathcal{D}_{i0i0}(z)
&=& -(d-1)\Bigg[\mathcal{D}(z^2)+\mathcal{D}_1(z^2)
\nn\\
& & \qquad\qquad\qquad 
+ z^2\frac{\partial\mathcal{D}_1(z^2)}{\partial z^2}\Bigg].
\eea
The leading-order contribution to the field strength correlator is given by the diagram shown in
Fig. \ref{figfscLO}. In $d = 4 - 2\epsilon$, the result is 
\begin{equation}
\mathcal{D}_1^{(0)}(z^2)=\mu^{2\epsilon}(N_c^2-1)\frac{\Gamma(2-\epsilon)}
{\pi^{2-\epsilon}(-z^2)^{2-\epsilon}}, \qquad 
\mathcal{D}^{(0)}(z^2)=0.
\end{equation}
Note that keeping $\epsilon \neq 0$ in the chromoelectric correlator 
provides a regularization for the integral over $t$ in Eq. (\ref{GpNRQCD}).

We now insert {\it (i)}, {\it (ii)} and {\it (iii)} into Eq. (\ref{GpNRQCD}).
Since at the ultrasoft scale $t(V_o-V_s) \sim 1$,  
the integral in $t$ is performed without expanding the exponential and making
use of 
\be
\int_0^{\infty}dt\,t^n\, e^{-at}=\frac{\Gamma(n+1)}{a^{n+1}}.
\label{intexp}
\ee
The final result reads:
\bea
\delta_{\rm US}&=&
C_F\frac{C_A^3}{24}\frac{1}{r}\frac{\als(\mu)}{\pi}\als^3(1/r)
\left(\frac{1}{\hat\epsilon}- 2 \log \frac{V_o-V_s}{\mu} +\frac{5}{3}
-2 \log 2\right),
\label{uslo}
\eea
where $\displaystyle \frac{1}{\hat\epsilon} = \frac{1}{\epsilon} -\gamma_E+\log (4 \pi)$.
Note that the $\als$ coming from the potential is evaluated at the soft
scale $1/r$, while the $\als$ coming from the ultrasoft coupling is evaluated
at the scale $\mu$. This will become relevant in the next section.
The ultraviolet divergence in (\ref{uslo}) can be reabsorbed 
by a renormalization of the potential. In the $\MS$ scheme, in coordinate space, we have: 
\be
V_s(r;\mu)\rightarrow Z V_s(r;\mu), \quad\quad Z
=1+\frac{C_A^3}{24}\frac{\als (\mu)}{\pi}\als^2(1/r)
\frac{1}{\hat\epsilon}.
\label{sub}
\ee
Since the static energy is $\mu$ independent, 
from the calculation above we infer that the logarithmic
contribution to $V_s(r;\mu)$ at order $\als^4$  must be 
\be
\delta V_s(r;\mu)=
-C_F {C_A^3\over 12} {1\over r}{\als(\mu)\over \pi} \als^3(1/r) \log ({r\mu}),
\ee
which added to the renormalized $\delta_{\rm US}$ contribution (from (\ref{uslo})) 
gives the $\log \als$ term  displayed 
in the third line of Eq. (\ref{E0stat}). This term was first calculated in
\cite{Brambilla:1999qa}, where the cancellation between the IR cut-off
of $V_s(r;\mu)$  and the UV cut-off of the pNRQCD expression was checked
explicitly by calculating the relevant Feynman diagrams in the Wilson loop. 

A comment is in order concerning the scheme dependence of the calculation of
$\delta_{\rm US}$. This is not important if we are only interested in the
logarithmic contribution, but it is if we wish eventually to combine our
result with a (yet to be done) calculation of $V_s (r; \mu )$ at N$^3$LO and
get the non-logarithmic pieces of the static energy right. We will assume that
such a calculation will be done in momentum space and that dimensional regularization 
and the $\MS$ scheme will be used to renormalize the UV divergences, like
in the N$^2$LO calculation \cite{Peter:1996ig,Peter:1997me,Schroder:1998vy}. The
result will still be IR divergent when $d\rightarrow 4$, and the question is
how one should proceed in order to combine that result with ours in a
consistent way.\footnote{Note that the $\MS$ subtraction of
(\ref{sub}) in coordinate space is not equivalent to the 
$\MS$ subtraction in momentum space.} We propose to convert the (UV
renormalized) momentum-space potential to coordinate space 
(in $d$ dimensions) in that calculation, and together  to
use $d$-dimensional expressions for all the objects in our calculation, namely
also for $V_s(r;\mu )$ and $V_o(r;\mu )$ ($V_A(r;\mu)$ remains the same
in $d$ dimensions). This guarantees that the IR behavior of the regulated
effective theory is exactly the same as the one of the fundamental theory.  
Had we expanded $V_o-V_s$ in Eq. (\ref{GpNRQCD}) we would have obtained zero,
which means that the UV divergences, which remain after renormalization by the $\MS$ QCD
counterterms (and by that of the color octet field wave function) in the effective theory, cancel exactly the IR divergences. 
Therefore, as a consequence of the fact that the IR behavior of the regulated
effective theory is the same as the one of the fundamental one, 
the UV divergences in (\ref{uslo}) cancel exactly the IR divergences in 
$V_s(r;\mu)$, the $\mu$ dependence 
disappears, and the non-logarithmic pieces are correctly calculated. This procedure would be
analogous to the one employed in \cite{Kniehl:2002br}.   
Alternatively, one could use $\MS$ for the IR divergences of $V_s(r;\mu)$ in
momentum space, work out the momentum space expressions for the $d$-dimensional
version of (\ref{uslo}) and make the $\MS$ UV subtraction accordingly.

In the following section, we will use the same procedure employed here to
obtain the next-to-leading IR logarithmic dependence of the static
potential. That is the logarithmic $\als^5$ contribution to the potential, 
which is part of the N$^4$LO contribution.

\section{Fourth-order logarithmic correction}
\label{sec3}
Equation (\ref{GpNRQCD}) does not rely on an expansion in $\als$, therefore 
it also provides NLO contributions to $\delta_{\rm US}$. In fact, as we argue next, it 
provides the full contribution to this order.

In principle, we may have diagrams with more insertions of the operators in (\ref{pNRQCD})
and diagrams with operators of higher order in the multipole expansion that 
contribute to $\delta_{\rm US}$ at NLO. Concerning the former, for symmetry
reasons we need at least two more operator insertions, which implies a
suppression of $\als^3$ with respect to the leading-order $\delta_{\rm US}$. 
Concerning the latter, operators of higher order in the multipole expansion 
may be found in \cite{Brambilla:2002nu,Brambilla:2003nt}. Their contributions are suppressed 
by $\als^2$ with respect to the leading-order $\delta_{\rm US}$.  To see
this just  recall that  the ultrasoft fields (and derivatives acting on them)
must be counted as $E_0 (r) \sim \als /r$. Then, any insertion of the kind
$\displaystyle \int dt \, \mathbf{r\cdot E}$ implies an $\als$ suppression 
(with an extra $\als$ suppression for any  ${\bf r}^i{\bf D}^j$ acting on the
chromoelectric field).  For a given diagram, additional suppressions may
appear due to the coupling constants in front of the chromoelectric fields.

The NLO contribution to $\delta_{\rm US}$ is then provided by Eq.
(\ref{GpNRQCD}) evaluated at relative order $\als$.
Since the dependence in $\als$ enters through $V_A$, $V_s$, $V_o$ and
the chromoelectric correlator, we need the $\mathcal{O}(\als)$ corrections to all these
quantities. These will be given in the following two sections.  Finally, in
Sec. \ref{subseccal4o}, we will obtain the fourth-order logarithmic
correction to the potential.

\subsection{$\mathcal{O}(\als)$  corrections of $V_A$, $V_s$ and $V_o$}
The $\mathcal{O}(\als)$ corrections to $V_s$ and $V_o$ are well known. 
In particular, we have
\be
V_o - V_s =  
\frac{C_A}{2}\frac{1}{r}\als(1/r)\left[1+\left(a_1+2\gamma_E\beta_0\right)\frac{\als(1/r)}{4\pi}\right].
\ee

\begin{figure}
\centering
\includegraphics{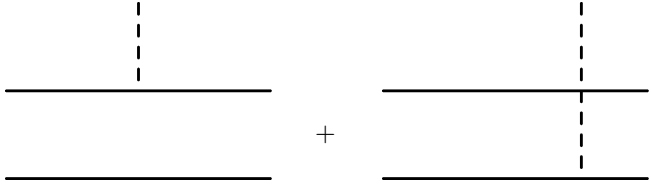}
\caption[Leading order matching of $V_A$]{Static QCD diagrams for the leading-order 
matching of $V_A$. The solid lines stand for a static quark and antiquark, 
the dashed line for a longitudinal gluon.}
\label{figVALO}
\end{figure}

\begin{figure}
\centering
\includegraphics{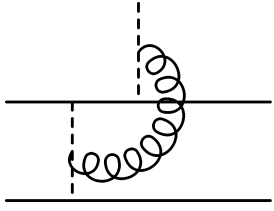}
\caption[Next-to-next-to-leading  order matching of $V_A$]{
Example of static QCD diagram that contributes to  
the next-to-next-to-leading order matching of $V_A$.}\label{figVANLO}
\end{figure}

The matching coefficient $V_A$ can be obtained by matching static QCD to pNRQCD at order
$r$ in the multipole expansion. At leading order in $\als$, we have
to calculate the diagrams shown in Fig. \ref{figVALO}. They give the tree
level result $V_A=1$. One may naively expect the first correction to be
$\mathcal{O}(\als)$, but it is not.\footnote{
The vanishing of the anomalous dimension of $V_A$ at one loop has been 
observed in \cite{Pineda:2000gz}.} 
This becomes clear if we perform the calculation in dimensional regularization and in 
Coulomb gauge\footnote{The potentials are independent on the gauge used in the matching. 
Therefore, we can use the most convenient one to do the computation.}. Indeed, the 
diagrams that we can draw at $\mathcal{O}(\als)$ correspond either to self-energy corrections 
or to iterations of the Coulomb potential, which are identical in the
effective theory and hence do not contribute to the matching. 
Then, the first non-vanishing correction to the tree level result may possibly come from 
diagrams like the one in Fig. \ref{figVANLO}, which is $\mathcal{O}(\als^2)$
and, therefore, unimportant here.


\begin{figure}
\centering
\includegraphics{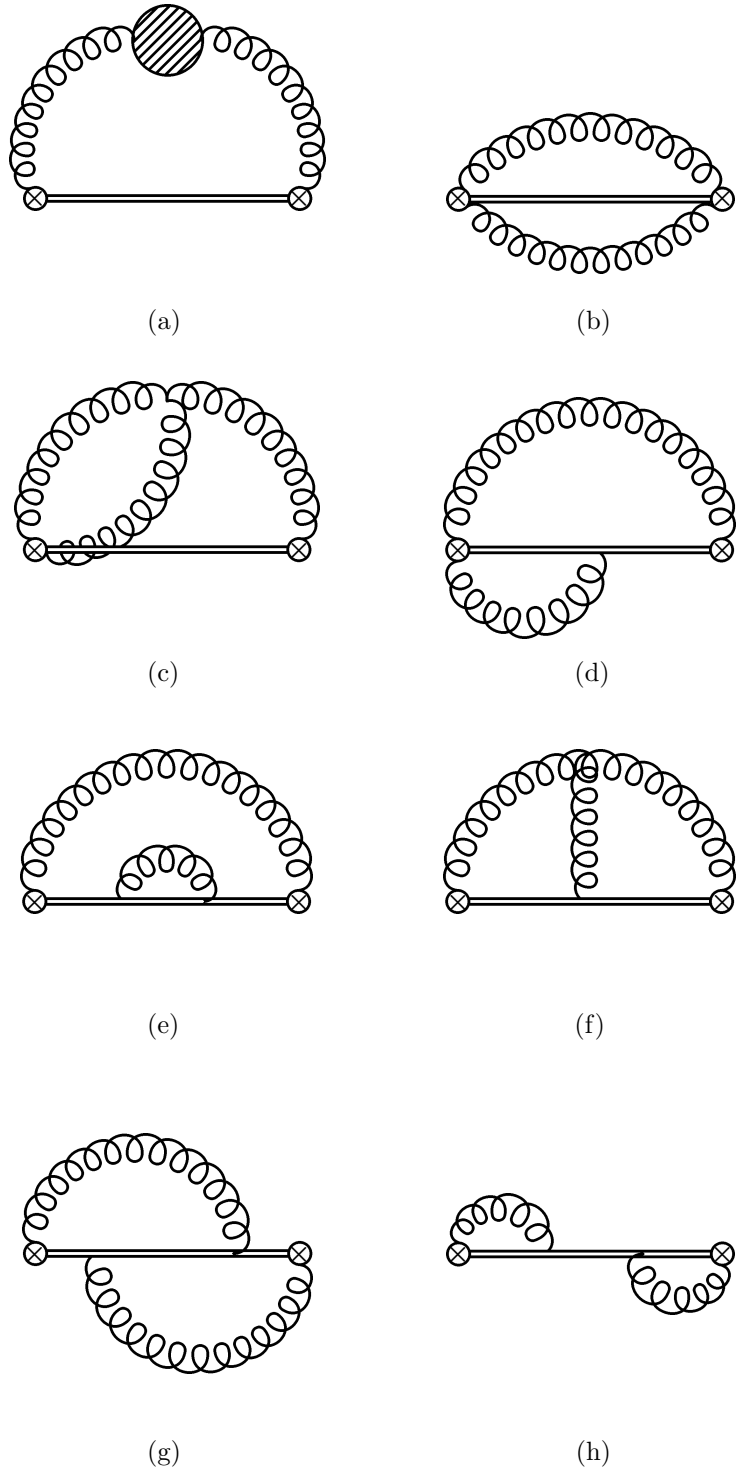}
\caption[Field strength correlator at next-to-leading order]{Next-to-leading
order contributions to the field strength correlator. The gluonic string is
represented by a double line. The shaded blob represents the insertion of the
one-loop gluon self-energy. Symmetric graphs are understood for (c) and
(d).}\label{figfscNLO}
\end{figure}

\subsection{$\mathcal{O}(\als)$ correction of the field strength correlator}
The $\mathcal{O}(\als)$ correction to the QCD field strength correlator was calculated in \cite{Eidemuller:1997bb}. 
It is given by the diagrams in Fig. \ref{figfscNLO}. Here we need the expression in $d$
dimensions because in (\ref{GpNRQCD}) the integral over $t$ 
is singular. The $d$-dimensional result for the $\als$ correction is \cite{pcJamin}
\begin{eqnarray}
\mathcal{D}^{(1)}(z^2) & = & N_c(N_c^2-1)\frac{\als(\mu)}{\pi}
\frac{\mu^{4\epsilon}}{4\pi^{2-2\epsilon}}\Gamma^2(1-\epsilon)\left(\frac{1}{z^2}\right)^{2-2\epsilon}g(\epsilon),\\
\mathcal{D}_1^{(1)}(z^2) & = & N_c(N_c^2-1)\frac{\als(\mu)}{\pi}
\frac{\mu^{4\epsilon}}{4\pi^{2-2\epsilon}}\Gamma^2(1-\epsilon)\left(\frac{1}{z^2}\right)^{2-2\epsilon}g_1(\epsilon),
\end{eqnarray}
with
\begin{eqnarray}
g(\epsilon) &=& 
\frac{-3 + 8\epsilon -6 \epsilon^2 + 2 \epsilon ^3}{\epsilon  \left(3 -5 \epsilon + 2
  \epsilon ^2\right)}
+ 2 \epsilon \frac{B(-1 + 2 \epsilon,-2 + 2 \epsilon)}{3 -2 \epsilon},
\\
g_1(\epsilon) &=& 
\frac{6 -18 \epsilon + 17 \epsilon ^2 -6 \epsilon ^3}
{\epsilon ^2 \left(3 -5 \epsilon + 2 \epsilon ^2\right)}
- 2 (1-\epsilon+\epsilon^2) \frac{B(-1 + 2 \epsilon,-2  + 2 \epsilon)}{\epsilon  (3 - 2 \epsilon)}
\nn\\
&& 
- 4 \, \frac{T_F n_f}{N_c} \, \frac{1-\epsilon}{\epsilon (3 - 2 \epsilon)},
\end{eqnarray}

where

$B(u,v)=\Gamma (u)\Gamma (v)/\Gamma (u+v)$. 

\begin{figure}
\centering
\includegraphics{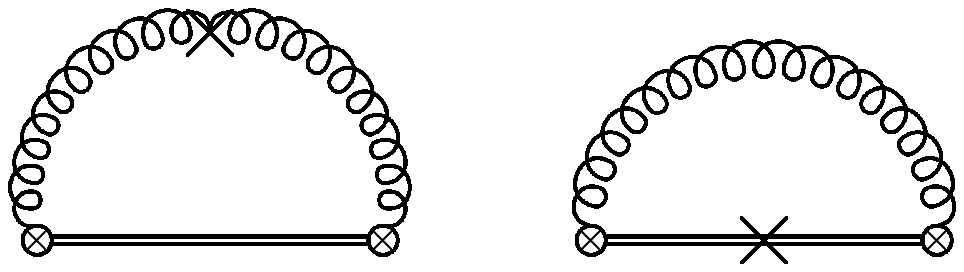}
\caption[$\mathcal{O}(\als)$ counterterm diagrams for the chromoelectric
  correlator]{$\mathcal{O}(\als)$ counterterm 
diagrams for the chromoelectric correlator. The gluonic string (which comes
  from the octet propagator) is represented by a double line.}
\label{figctEE}
\end{figure}

Since the external points $x$ and $y$ are fixed, 
the divergences that we encounter in $\mathcal{D}_{i0i0}$ coming from the
expressions above should cancel against 
the vertex and gluon and octet field propagator counterterms. The
counterterm for the vertex is zero, since, as seen in the previous
section, the first correction to $V_A$ is of order $\als^2$. 
The counterterm for the gluon propagator is the usual
one in QCD. The counterterm for the octet propagator coincides with the
counterterm for the quark propagator in the Heavy Quark Effective Theory \cite{Neubert:1993mb}
but with the quark in the adjoint representation. We can represent the counterterm
contributions by the diagrams of Fig. \ref{figctEE}. We have checked that:
{\it (i)} the divergence coming from the first diagram in Fig. \ref{figfscNLO} is
canceled by the counterterm of the gluon propagator, {\it (ii)} the diagram (b) 
of Fig. \ref{figfscNLO} does not give a divergent contribution (as one would expect from the
fact that the gluons are attached to the external fixed points only) 
and {\it (iii)}  when we sum the remaining diagrams the divergence that we obtain is exactly
canceled by the counterterm of the octet propagator.
In the $\MS$ scheme, at $\mathcal{O}(\als)$, the contributions of the counterterms are given by
\begin{eqnarray}
\mathcal{D}^{\mathrm{c.t.}}(z^2) & = & 0\\
\mathcal{D}^{\mathrm{c.t.}}_1(z^2) & = &
N_c(N_c^2-1)\frac{\als(\mu)}{\pi}\frac{\mu^{2\epsilon}}
{4\pi^{2-\epsilon}}\Gamma(2-\epsilon)\frac{1}{(-z^2)^{2-\epsilon}}
\frac{1}{\hat\epsilon} 
\nn\\
& & \qquad\qquad\qquad\qquad\qquad\qquad 
\times \Bigg(-2 -\frac{5}{3}+\frac{4}{3}T_F\frac{n_f}{N_c}\Bigg)
\end{eqnarray}
where in the brackets we have kept separated the $-2$ 
coming from the octet propagator counterterm 
from the $\displaystyle -\frac{5}{3}+\frac{4}{3}T_F\frac{n_f}{N_c}$ coming from the gluon propagator
one. The renormalized $d$-dimensional result for the $\als$ correction to the chromoelectric correlator is
\bea
\mathcal{D}_{i0i0}^{(1)} &=&
-(3-2\epsilon)\left[\mathcal{D}^{(1)}(z^2)+(-1+2\epsilon)\mathcal{D}_1^{(1)}(z^2)
\right.
\nn\\
&& \left. \qquad \qquad \qquad \qquad \qquad 
+\mathcal{D}^{\mathrm{c.t.}}(z^2)+(-1+\epsilon)\mathcal{D}_1^{\mathrm{c.t.}}(z^2)\right],
\label{EENLO}
\eea
which, indeed, is finite for  $\epsilon \to 0$.

\subsection{Calculation of the fourth-order logarithmic correction}\label{subseccal4o}
The results of the two preceding sections provide all the necessary ingredients to compute 
$\delta_{\rm US}$ at NLO. Let us split $\delta_{\rm US}$  as follows
\be
\delta_{\rm US}=
\mathcal{G}^{(r^2)}_{\langle EE\rangle \vert_{\mathcal{O}(\als)}}
+\mathcal{G}^{(r^2)}_{V_o-V_s\vert_{\mathcal{O}(\als)}}
+\mathcal{G}^{(r^2)}_{\mu\to 1/r\vert_{\mathcal{O}(\als)}},
\label{deltaUS}
\ee
where the first and second terms stand for the $\als$ corrections 
to the field strength correlator and to the potentials respectively, 
and the last term accounts for the contribution induced by a change of scale in the N$^3$LO calculation. 

First, we shall consider the contribution (\ref{EENLO}) to the field strength correlator. 
After integration over $t$, which can be done using Eq. (\ref{intexp}), 
we obtain 
\bea
&& \hspace{-8mm}
\mathcal{G}^{(r^2)}_{\langle EE \rangle\vert_{\mathcal{O}(\als)}}=
\left(\frac{\als(\mu)}{\pi}\right)^2\als^3(1/r) \, C_F \frac{C_A^3}{8}\frac{1}{r}
\nn\\
&& 
\hspace{15mm}
\times \left[
 \frac{A}{\hat\epsilon^2} 
+ \frac{B}{\hat\epsilon} 
+C_1\log^2\frac{V_o-V_s}{\mu}
+C_2\log\frac{V_o-V_s}{\mu}
+D\right]\!,
\label{coEE}
\eea
with
\begin{eqnarray}
A & = & \frac{1}{24} \left(\frac{4 \, T_F n_f}{3}-\frac{11}{3}C_A\right),\\
B & = & \frac{1}{108} \left[-10 \, T_F n_f+C_A\left(6 \pi ^2+47\right)\right], \\
C_1 & = & \frac{1}{6} \left(-\frac{4 \, T_F n_f}{3}+\frac{11}{3}C_A\right), \\
C_2 & = & \frac{1}{54} \left[4 \, T_F n_f\left(10 - 6 \log 2 \right)
+C_A\left( -149 + 66 \log 2 -12 \pi ^2\right)\right], \\
D &=&  \frac{1}{9}\left[
T_Fn_f \left( - \frac{67}{9} + \frac{5}{6} \gamma_E + 5 \log 2 -2 \log^2 2 
- \frac{5}{6} \log \pi - \frac{\pi^2}{3} \right) \right.
\nn\\
&& 
+ C_A \Bigg(\frac{1241}{36} - \frac{47}{12} \gamma_E 
- 17 \log 2 + \frac{11}{2} \log^2 2 + \frac{47}{12} \log \pi -12 \zeta(3)
\nn\\
&&  \quad
\left.
+  \frac{9}{4}\pi^2  - \frac{\gamma_E}{2}\pi^2  - \pi^2 \log 2 + \frac{\pi^2}{2} \log \pi \Bigg)
\right].
\end{eqnarray}

Next, we display the contribution that we obtain if in (\ref{GpNRQCD})
we use the leading-order expression for the chromoelectric correlator but the 
${\mathcal O}(\als)$ correction for $V_o-V_s$: 
\bea
\mathcal{G}^{(r^2)}_{V_o-V_s\vert_{\mathcal{O}(\als)}} & = & 
\frac{\als(\mu)}{\pi}\frac{\als^4(1/r)}{\pi}C_F\frac{C_A^3}{16}\frac{1}{r}\left(a_1+2\gamma_E\beta_0\right)
\nn\\
&& \qquad\qquad 
\times\left[\frac{1}{2\,\hat\epsilon} 
-\log\left(\frac{V_o-V_s}{\mu}\right)+\frac{5}{6}-\log2\right].
\label{coprop}
\eea

The ultraviolet divergences in the expressions (\ref{coEE}) and (\ref{coprop})
come from the integration over time in (\ref{GpNRQCD}).
They can be absorbed by a renormalization of the potential, analogous to (\ref{sub}).

Finally, we obtain another contribution if in the renormalized version of (\ref{uslo}) we
change $\als(\mu)$ to $\als(1/r)$ (we want all $\als$ evaluated at the scale $1/r$):
\bea
 \mathcal{G}^{(r^2)}_{\mu\to 1/r\vert_{\mathcal{O}(\als)}}&=&
\frac{\als^5(1/r)}{\pi^2}C_F\frac{C_A^3}{24} \frac{1}{r}\beta_0 \log(r\mu)
\nn\\
&& \qquad\qquad 
\times \left[\log\left(\frac{V_o-V_s}{\mu}\right)+\log2-\frac{5}{6}\right].
\label{cormu}
\eea

Adding up the renormalized versions of (\ref{coEE}) and (\ref{coprop}) and
 equation (\ref{cormu}), we obtain the contribution of $\delta_{\rm US}$ to
$E_0 (r)$ at order $\als^5$. The complete calculation of $E_0 (r)$ at this order requires the
knowledge of $V_s (r;\mu )$ at the same order. However, to obtain the
terms proportional to $\log \als$ it is enough to enforce $E_0 (r)$ to 
be independent of the factorization scale $\mu$.  This constrains the 
terms $\als^5\log^2 r\mu$ and $\als^5\log r\mu$ of the singlet static potential to be
\bea 
&& \delta V_s (r;\mu )
= -\frac{C_F\als(1/r)}{r}\left(\frac{\als(1/r)}{4\pi}\right)^4
\nn\\
&& \qquad 
\times 
\left\{\frac{16\pi^2}{3}C_A^3\left(-\frac{11}{3}C_A+\frac{4}{3}T_Fn_f\right)\log^2r\mu
\right.
\nn\\ 
&& \qquad\quad
+ 16\pi^2C_A^3\left[a_1+2\gamma_E\beta_0 
-\frac{20}{27} T_F n_f
\left.
+ C_A\left(\frac{94}{27}+\frac{4}{9}\pi^2\right)\right]
\log r\mu\right\}. 
\label{V0stat}
\eea
Summing (\ref{V0stat}) with (\ref{deltaUS}) provides the coefficients $a_4^{L2}$
and $a_4^{L}$ of the static energy $E_0(r)$ given in Eqs. (\ref{a4L2}) and
(\ref{a4L}) respectively.

Note that: {\it (i)}  in order to cancel the $\mu$ dependence of the two  
double logarithms in $\delta_{\rm US}$, $\log(r\mu)\log((V_o-V_s)/\mu)$ and $\log^2((V_o-V_s)/\mu)$,  
against the single double logarithm in $\delta V_s$, $\log^2r\mu$,  
the coefficient of $\log(r\mu)\log((V_o-V_s)/\mu)$ must be twice the one  
of $\log^2((V_o-V_s)/\mu)$. {\it (ii)} The coefficient of the double logarithm 
$\log^2r\mu$  in $\delta V_s$ should coincide with the one obtained expanding  
the renorma\-lization group improved static potential of \cite{Pineda:2000gz}. 
{\it (iii)} The coefficients $a_4^{L2}$ and $a_4^{L}$ must be renormalization scheme 
independent\footnote{For a calculation of $\delta V_s(r;\mu)$ in the
subtraction scheme $1/\epsilon -\gamma_E + \log \pi$
we refer to \cite{GarciaPHD}.}. We have explicitly checked that our result satisfies these
requirements.

\section{Conclusions}
\label{sec4}
We have calculated the ultrasoft contribution to the QCD static energy 
of a quark-antiquark pair at order $\als^5$. 
This is sufficient to obtain the logarithmic contribution to the
pNRQCD singlet static potential at N$^4$LO, which, in turn, provides the 
$\als^5 \log^2 \als/r$  and $\als^5 \log \als/r$ 
terms of the static energy of a quark-antiquark pair 
at distance $r$.  The calculation heavily relies
on effective field theory techniques and uses the result of
Ref. \cite{Eidemuller:1997bb} as a key ingredient.

Possible applications of the result include precision comparisons with lattice data, 
heavy quarkonium spectra and  $t$-$\bar t$ production near threshold.

At short distances, the perturbative expression of the QCD static energy 
has been compared with lattice data at N$^2$LO in \cite{Sumino:2001eh,Necco:2001xg}
and at N$^2$LL in \cite{Pineda:2002se}. Our analysis provides a key ingredient 
for a N$^3$LL analysis.

Starting from the N$^3$LO in $\als$, the quarkonium mass becomes sensitive 
to the ultrasoft scale, if the ultrasoft scale is assumed to be much larger
than $\lQ$ \cite{Kniehl:1999ud,Brambilla:1999xj,Kniehl:2002br,Penin:2002zv}.
Our result also provides an important ingredient for the calculation of the quarkonium mass at N$^3$LL accuracy. 

Top-quark pair production near threshold, which will become an important
production process at the ILC, is presently known at N$^2$LO  \cite{Hoang:2000yr}. 
The cross section at N$^2$LL (see e.g. \cite{Hoang:2003ns,Pineda:2006ri})
and at N$^3$LO (see e.g. \cite{Beneke:2005hg,Penin:2005eu}) is computed 
presently by several different groups. Our result will contribute to the 
cross section at N$^3$LL. The third-order renormalization group improved 
expression will be needed to resum logarithms potentially as large as 
the N$^3$LO and reduce the scale dependence of the cross section.

{\bf Acknowledgments}

We are grateful to Matthias Jamin for making us available the details 
of the calculations reported in Ref. \cite{Eidemuller:1997bb}. 
N.B. and A.V. thank Yu-Qi Chen, Carlo Ewerz and Yu Jia for discussions. Part of this work has been carried out at
ECT*, Trento, in August 2006, during the program ``Heavy
quarkonium and related heavy quark systems''. 
We acknowledge financial support from ``Azioni Integrate Italia-Spagna 2004 (IT1824)/Acciones Integradas 
Espa\~na-Italia (HI2003-0362)'', and from the cooperation agreement INFN05-04
(MEC-INFN). X.G.T. and J.S. are also supported by MEC (Spain) grant CYT FPA
2004-04582-C02-01, the CIRIT (Catalonia) grant 2005SGR00564 and the network
Euridice (EU) HPRN-CT2002-00311. X.G.T. acknowledges financial support 
from the DURSI of the Generalitat de Catalunya and the Fons Social Europeu. 
The work of X.G.T. was also supported in part by the U.S. Department of
Energy, Division of High Energy Physics, under contract W-31-109-ENG-38. 
A.V. acknowledges the financial support obtained inside the Italian
MIUR program  ``incentivazione alla mobilit\`a di studiosi stranieri e
italiani residenti all'estero''.

\end{document}